\documentclass{aipproc}
\layoutstyle{8x11single}

\begin{document}

\title{Planck Low Frequency Instrument: Beam Patterns}

\classification{43.35.Ei, 78.60.Mq}

\author{M.Sandri}{
  address={TeSRE -- CNR, 40122 Bologna, Italy},
  email={sandri@tesre.bo.cnr.it},
}
\author{M.Bersanelli}{
  address={Universit\`a di Milano, 20133 Milano, Italy},
  email={marco@ifctr.mi.cnr.it},
}
\author{C.Burigana}{
  address={TeSRE -- CNR, 40122 Bologna, Italy},
  email={burigana@tesre.bo.cnr.it},
}
\author{C.R.Butler}{
  address={TeSRE -- CNR, 40122 Bologna, Italy},
  email={butler@tesre.bo.cnr.it},
}
\author{M.Malaspina}{
  address={TeSRE -- CNR, 40122 Bologna, Italy},
  email={mala@tesre.bo.cnr.it},
}
\author{N.Mandolesi}{
  address={TeSRE -- CNR, 40122 Bologna, Italy},
  email={reno@tesre.bo.cnr.it},
}
\author{A.Mennella}{
  address={IFCTR -- CNR, 20133 Milano, Italy}, 
  email={daniele@ifctr.mi.cnr.it},
}
\author{G.Morgante}{
  address={TeSRE -- CNR, 40122 Bologna, Italy},
  email={morgante@tesre.bo.cnr.it},
}
\author{L.Terenzi}{
  address={TeSRE -- CNR, 40122 Bologna, Italy},
  email={lteren@tesre.bo.cnr.it},
}
\author{L.Valenziano}{
  address={TeSRE -- CNR, 40122 Bologna, Italy},
  email={valenziano@tesre.bo.cnr.it},
}
\author{F.Villa}{
  address={TeSRE -- CNR, 40122 Bologna, Italy},
  email={villa@tesre.bo.cnr.it},
}

\copyrightyear{2001}

\begin{abstract}
The Low Frequency Instrument on board the Planck satellite is coupled to the Planck 1.5 meter off-axis dual 
reflector telescope by an array of 27 corrugated feed horns operating at 30, 44, 70, and  100 GHz. 
We briefly present here a detailed study of the optical interface devoted to optimize the angular resolution 
(10 arcmin at 100 GHz as a goal) and at the same time to minimize all the systematics coming from the sidelobes 
of the radiation pattern. 
Through optical simulations, we provide shapes, locations on the sky, angular resolutions, and polarization 
properties of each beam.\hspace{0.2cm}{\it (On behalf of LFI Collaboration)}
\end{abstract}

\maketitle

\section{Introduction}
The Planck Telescope is designed as an off-axis tilted system offering the advantage of an unblocked aperture. 
The baseline configuration has been selected among thirty different optical designs. 
The current configuration has been obtained by optimizing the telescope performance for a set of equally 
distributed in frequency (from 30 to 857 GHz) and space (within the Focal Plane Box) representative feed horns. 
Both mirrors have an ellipsoidal shape (Aplanatic Configuration). 
The conical constants, the focal lenght, the tilting, and the decenter of the mirrors have been combined 
to reduce the main beam aberrations, the curvature of the focal surface, and the spillover as well.

The Low Frequency Instrument (LFI) is one of the two instruments onboard the Planck satellite 
\cite{SCI-PT-RS-07024} that shares the focal region of the telescope. 
It is an array of 54 HEMT-based pseudo-correlation receivers coupled to the Telescope by 27 dual 
profiled corrugated feed horns. 
The baseline layout of the LFI Focal Plane Unit (FPU) foresees the 27 feed horns located around the 
High Frequency Instrument (HFI): 16 feed horns at 100 GHz, 6 at 70 GHz, 3 at 44 GHz and 2 at 30 GHz.
The LFI team has carried out an exhaustive study of the LFI optical interface devoted to optimize the 
angular resolution (12 arcmin at 100 GHz as a requirement, 10 arcmin at 100 GHz as a goal) and at the 
same time to minimize all the systematics coming from the side lobes of the radiation pattern. 

\section{Optical Simulations}
Different techniques can be applied to predict the radiation pattern: Geometrical Optics (GO), 
Geometrical Theory of Diffraction (GTD), Physical Optics (PO) and Physical Theory of Diffraction (PTD). 
The simulations have been performed considering each feed (Gaussian, X-- axis polarized \cite{Vilas:2000}) 
as a source and computing the pattern scattered by both reflectors on the far field. 
The GO/GTD methods have been used to model the sub reflector, while the main reflector has been 
modelized using the PO. 

\subsection{Angular Resolution vs Edge Taper}
The Angular Resolution (expressed here in term of Full Width Half Maximum, FWHM) of the beam on the sky 
depends on the illumination of the primary mirror. 
The illumination can be represented by the Edge Taper, defined as the ratio of the power per unit area 
incident on the centre of the mirror to that incident on the edge. 
Decreasing the edge taper has a positive impact on the angular resolution but lower is the edge taper 
and higher is the side lobes level since they are largely due by diffraction and scattering from the 
reflector edges. 
The trade-off between the Angular Resolution (which impacts the ability to reconstruct the anisotropy 
power spectrum of the Cosmic Microwave Background Radiation at high multipoles \cite{Reno:1991}) and the Edge Taper 
(which impacts the systematics of the detected signal from receivers \cite{Carlo:2001}) has been carried out for some LFI 
feed horns \cite{IR308:2001}. 
The dependence of the angular resolution improvement on the edge taper degradation is almost linear until 
a certain edge taper is reached, when increasing the illumination on the primary mirror doesn't involve a further 
angular resolution betterment. 
The latter because a strong illumination of the mirrors increases the aberrations on the main beam.
Obviously, the amount of the improvement depends on the feed horn location, since the
primary mirror is illuminated in a different way.

\subsection{Field Distribution on Primary Mirror}
The amplitude field incident on the main reflector has been computed for each LFI feed horn. 
The model of the feed we used is a X-- axis polarized gaussian horn with an edge taper of  
30 dB at 22$^\circ$ of angle \cite{Vilas:2000}. 
The X-- axis (Y-- axis) of the contour plot lies on the symmetry (asymmetry) plane of the telescope. 
The +Z direction corresponds to the main beam peak direction and the (X,Y,Z) is a standard
cartesian frame. 
As a conseguence, the top edge of the main reflector is at X $\simeq$ 750 mm and Y$\simeq$ 0 mm on the 
contour plot coordinate system. 
We used the Geometrical Optics (GO) and the Geometrical Theory of Diffraction (GTD) on the sub 
reflector to calculate the total amplitude of the field incident on the surface of the primary 
mirror, in the reference system of the main beam.

\begin{figure}[!h]
\begin{tabular}{ccc}
  \hspace{0cm}\includegraphics[height=.23\textheight]{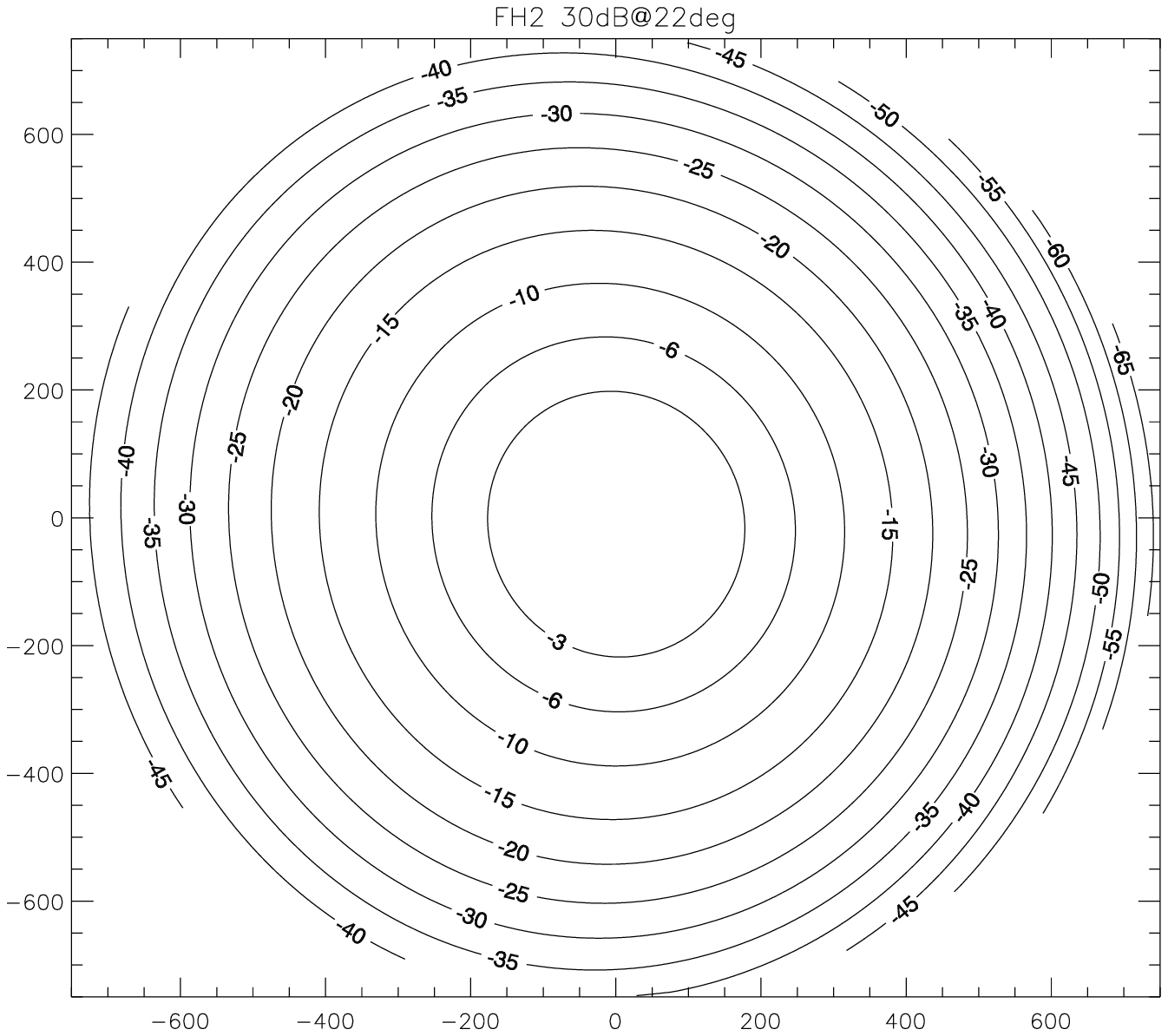}
&
  \hspace{-1.5cm}\includegraphics[height=.23\textheight]{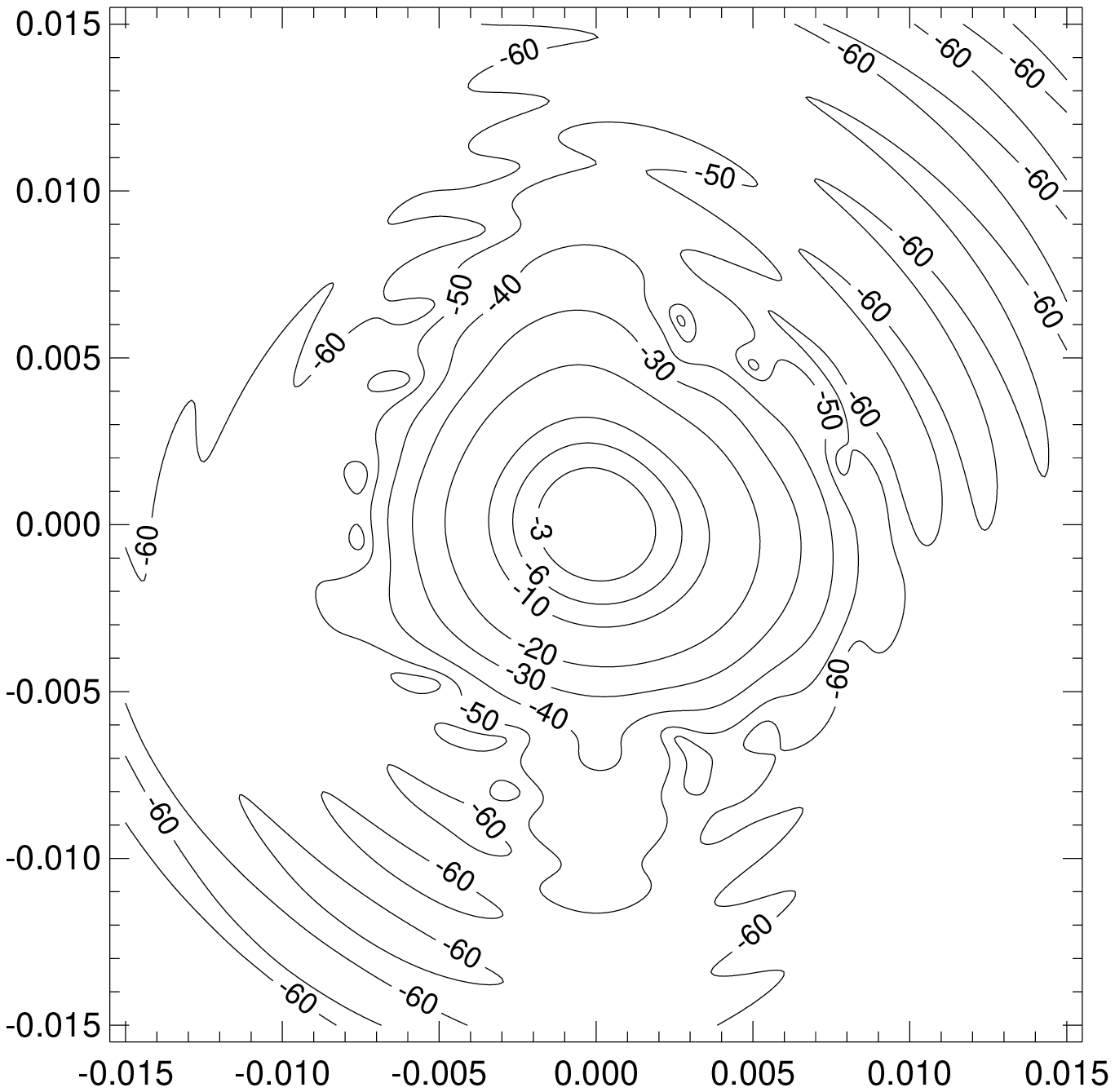}
&
  \hspace{-0.5cm}\includegraphics[height=.23\textheight]{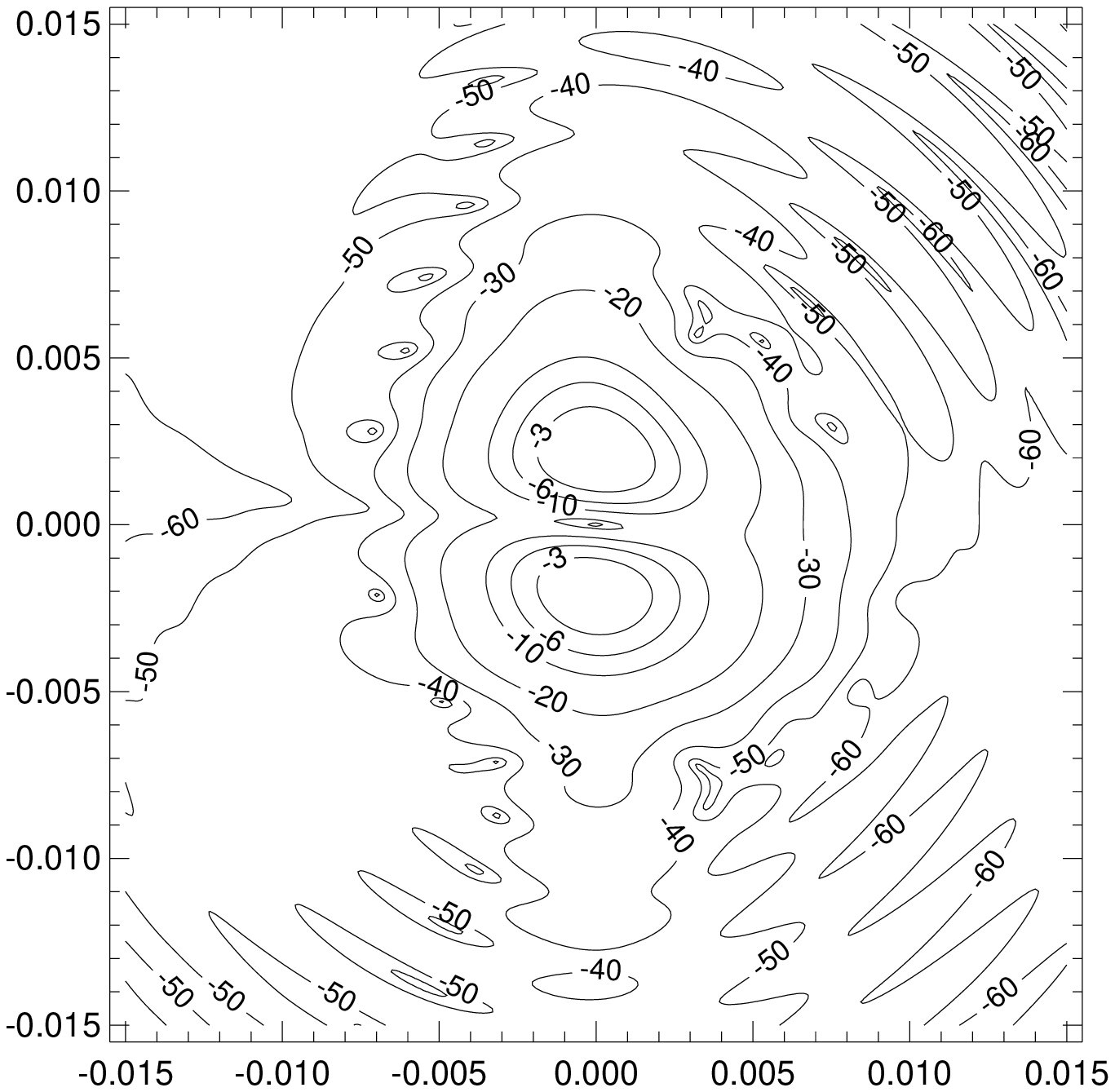}
\end{tabular}
\label{fig:inc}
\caption{Left: field distribution on the main reflector for the feed horn 2 at 100 GHz (X-- and
Y-- axis are in mm). 
Centre and right: UV plot of the copolar and crosspolar components of the feed horn 5 at 100 GHz after 
the polarization alignment.}
\end{figure}

The contour plots show that, as expected, the illumination of the primary mirror is roughly elliptical. 
As a consequence the field amplitude on the primary mirror rim is not constant 
(see Figure \ref{fig:inc}, left panel). 
The amplitudes of the field on the main reflector contour have been used to set the requirements on the 
edge taper values for all the LFI feed horn illuminations.
The field amplitude on the mirror contour is a function of the $\varphi$ angle ($E = E(\varphi)$), 
defined in the reference system of the main beam ($\varphi = 0$ in the direction of the top edge of the 
main reflector and has a counterclockwise direction). 

The edge taper of each feed, at a reference angle (22$^\circ$ or 24$^\circ$), has been chosen comparing 
the field amplitute, $E(\varphi)$, with that corresponding to a \textit{worst reference} case, 
$\tilde{E}(\varphi)$, for which a full straylight analysis has been performed, 
showing acceptable contamination levels from the galactic emission \cite{Carlo:2001}.
The edge taper correction of each feed horn, in order to assure a straylight rejection analogous 
to the reference case, was calculated by computing the lowest difference between the edge taper curve of 
each feed and the reference edge taper curve ($min|E(\varphi)- \tilde{E}(\varphi)|$). 
We decrease by this difference the horn's edge taper at the reference angle and the results are reported 
in Table \ref{tab:beams}. 
In this way, for each LFI feed horn, no single point on the main reflector rim has an edge taper value 
lower than that of the reference case \cite{IR309:2001}. 

\subsection{Main Beam Response}
The main beam power pattern has been simulated for each feed horn, with the edge taper value found 
out by the Edge Taper optimization procedure \cite{IR309:2001}. 
Edge Taper, Angular Resolution, Ellipticity and Directivity of each beam are shown in Table \ref{tab:beams}. 
These results (adopted as requirements for the current LFI baseline) can be considered as conservative. 
In fact, although the reported simulations have been done using a Gaussian model for each feed horn, preliminary 
studies show that dual profiled horns allow to reach a further improvement on the angular resolution. 
Moreover, a dedicated straylight study for each feed horn will be performed with the aim to relax the edge taper
and subsequently to improve the angular resolution more.  
\vspace{0.5cm}

\begin{table}[!h]
\begin{tabular}{cccccccc} 
\hline
\tablehead{1}{r}{b}{Feed} &
\tablehead{1}{c}{b}{Frequency} & 
\tablehead{1}{c}{b}{ET} &
\tablehead{1}{c}{b}{FWHM} &
\tablehead{1}{c}{b}{FWHM} &
\tablehead{1}{c}{b}{FWHM} &
\tablehead{1}{c}{b}{Ellipticity} &
\tablehead{1}{c}{b}{Directivity} \\
     & (GHz) & (dB @ $^\circ$) & (Min, arcmin) & (Max, arcmin) & (Ave, arcmin) &  & (dBi) \\
\hline  
27 -- 28 &  30	& 30.00 - 22.00 & 29.16 & 40.68 & 34.92 & 1.40	& 50.47 \\
25 -- 26 &  44	& 30.00 - 22.00 & 25.32 & 31.68 & 28.50 & 1.25	& 52.08 \\
24       &  44	& 30.00 - 22.00 & 20.04 & 27.84 & 23.94 & 1.39	& 53.77 \\
23 -- 18 &  70	& 20.50 - 22.00 & 12.36 & 15.72 & 14.04 & 1.27	& 58.37 \\
22 -- 19 &  70	& 23.10 - 22.00 & 12.60 & 16.08 & 14.34 & 1.28	& 58.23 \\
21 -- 20 &  70	& 25.00 - 22.00 & 12.60 & 16.44 & 14.52 & 1.30	& 58.07 \\
9 --  10 & 100	& 25.50 - 24.00 &  9.48 & 11.88 & 10.68 & 1.25	& 60.59 \\
8 -- 11	 & 100	& 26.80 - 24.00 & 10.08 & 12.36 & 11.22 & 1.23 & 60.16 \\
7 -- 12	 & 100	& 27.60 - 24.00 & 10.56 & 12.84 & 11.70 & 1.22	& 59.79 \\
6 -- 13	 & 100	& 27.70 - 24.00 & 11.16 & 13.20 & 12.18 & 1.18	& 59.52 \\
5 -- 14	 & 100	& 27.40 - 24.00 & 11.40 & 13.20 & 12.30 & 1.16	& 59.34 \\
4 -- 15	 & 100	& 28.10 - 24.00 & 11.88 & 13.68 & 12.78 & 1.15	& 59.16 \\
3 -- 16	 & 100	& 28.30 - 24.00 & 11.88 & 13.80 & 12.84 & 1.16	& 59.07 \\
2 -- 17	 & 100	& 28.10 - 24.00 & 12.12 & 13.80 & 12.96 & 1.14	& 59.05 \\
\hline
\end{tabular}
\caption{Main Beams characteristics.}
\label{tab:beams}
\end{table}

\vspace{0.5cm}

Polarization properties of the beams are mainly determined by the feed position and orientation in the 
Focal Plane Unit. The LFI polarization properties have been optimized rotating each feed horn about its 
axis, in order to obtain the right orientation of the polarization direction of each beam. 
The polarized radiation in the far field is fully described by giving the projection of the electric 
field vector in two mutually orthogonal directions. Far field amplitude radiation patterns reported 
here are given in the Co and Cross polar basis according to the Ludwig's 3rd definition.
The reference system in which each beam is computed has been rotated about their Z-- axis in order 
to find out the main polarization direction on the sky. 
When this condition is reached, a well defined minimum appears in the crosspolar component, in 
corrispondence to the maximum of the copolar component (see Figure \ref{fig:inc}).

\end{document}